# Elucidating the Influence of Native Defects on Electrical and Optical Properties in Semiconducting Oxides: An Experimental and Theoretical Investigation

Shashi Pandey[1*], Alok Shukla[2], Anurag Tripathi[1]

[1]*Department of Electrical Engineering IET Lucknow, Uttar Pradesh 226021, India*
[2]*Department of Physics, Indian Institute of Technology Bombay, Powai, Mumbai, Maharashtra 400076, India*

*Email: 2512@ietlucknow.ac.in\*, shukla@iitb.ac.in, anurag.tripathi@ietlucknow.ac.in*

**Abstract:**

Defect engineering using self-doping or creating vacancies in polycrystalline oxide based materials has profound influence on optical absorption, UV photo detection, and electrical switching. However, defects induced semiconducting oxide devices show enhancement in photo detection and photosensitivity, and hence still remain interesting in the field of optoelectronics. In this study, defect engineering in semiconducting oxide materials is discussed along with their roles in optoelectronic device-based applications. Theoretical investigations have been done for identifying defect states by performing first-principles electronic structure calculations, employing the density-functional theory. Particularly, in this work we have focused on probing the defect-induced changes in optical and electrical processes by means of experimental as well as computational investigations. Hence, systematic experimental measurements of optical absorption, electrical switching, charge density, and photocurrent in defect-rich semiconducting oxide samples have been performed. Our results suggest that defects can lead to enhancement of optoelectronic properties such as photocurrent, switching speed, optical absorption etc. for semiconducting oxide based devices.

## Introduction

Semiconducting oxides are well-known functional materials exhibiting interesting properties such as superconductivity [1], ferroelectricity [2,3], magnetism [4–6], and a wide range of dielectric properties [7,8]. Semiconducting metal oxides such as $TiO_2$ and $SrTiO_3$ are promising candidates for photo catalysis [9–11] because of their exceptional electronic [12] and optical properties [13], with high photochemical stability. It is well known that the oxygen vacancies (point defects) in semiconducting oxides [9,12,12,14,15] play a crucial role as far as changes in the electrical, optical and magnetic properties are concerned, as compared to their pristine phases. Defects are known to be primarily responsible for ionic transport phenomena in solids, because they provide easy hopping [16] sites (i.e. variable oxidation states of Ti) for ions or atoms to move through the crystal, accounting for both diffusion and ionic conductions. As a result of this, electrical and optical properties show considerable amount of variation as compared to the parent samples. The presence of extra defect-induced states in the visible region has the potential for enormous applications in enhancement of photo catalytic activity, solar cell industry [17,18], storage device application etc. Therefore, it is very important to study the effects of vacancies on optical properties of $TiO_2$ and $SrTiO_3$ [12,19]. In case of metal oxides with wide band gaps, the influence of defect states can be probed through valence band [10] and photoluminescence spectroscopy [20] etc. Pure $TiO_2$ and $SrTiO_3$, however, are not good candidates for photocatalysis [11,21], because they are active only under ultraviolet (UV) irradiation. But, after introducing vacancies in $TiO_2$ and $SrTiO_3$, their bandgap can be tuned towards the UV-VIS region [9,22,23]. Previously, in many studies magnetic properties due to vacancies at the various sites [24–26], has been probed. Furthermore, the self-doping [9] due to strontium, titanium and oxygen vacancies [12,15] has a profound influence on the optical properties of these materials. Oxygen vacancies ($O_v$) are pervasive point defects in transition-metal oxides (TMOs), and their presence impacts numerous physical properties such as electrical conductivity, and magnetism. Recently optical absorption spectroscopy has been widely used to probe the electronic structure of various materials [15,27,28] and the said spectroscopic technique is also found to be very useful to probe defects and disorder [29,30] present in a sample. Thus, it may be expected that 'fingerprints' of defect states present in the sample may also appear in optical absorption spectra [14] measured in the laboratory, as well computed using the first-principles density-functional theory (DFT). In recent times, Li [31] et al. showed the effects of point defects on the optical properties using DFT calculations, and observed that the point defects lead to extra states in the computed optical absorption spectra. Keeping this in view, we have also performed a DFT-based investigation, and studied the consequences of defects produced due to self-doping in transition metal oxides.

In this work, we have performed systematic experimental and first-principles theoretical studies of self-doped $SrTiO_3$ and $TiO_2$ to investigate the variations in their optical absorption, electrical switching, charge density, and photocurrent.

**Experimental Procedure:**

$TiO_2$ thin films were synthesized by chemical vapor deposition using precursor $Ti(H_3BNMe_2BH_3)_2$ with $H_2O$ as a co-reactant. While for $SrTiO_3$ precursor solutions were synthesized using semi-hydrated strontium acetate $[Sr(CH_3COO)_2.1/2H_2O]$ and titanium tetra-n-butoxide $[Ti(OC_4H_9)_4]$, Acetic acid and 2-methoxyethanol have been selected as primary solvents. All precursor solutions were deposited by Indium tin oxide substrate using spin-coating with spin of 3000 rpm for 30s. After spinning sample on the substrates, prepared films were kept in ambient air for 1 h to form gel films. All prepared films were then slowly heated with a rate of 2 $Kmin^{-1}$ to a temperature of $300^0C$ for the decomposition of residual organics. For structural confirmation of Anatase-$TiO_2$ and cubic $SrTiO_3$ Raman measurements (see Figure-1) were performed using Jobin-Yvon Horiba micro Raman spectrometer.

**Computational details**

All the calculations were performed within the framework of plane-wave density functional theory (PW-DFT)[32,33] implemented in Vienna ab-initio simulation package (VASP)[34,35]. The generalized gradient approximation (GGA+U) has been used for all calculations[36,37]. Calculations have been performed on $SrTiO_3$ and $TiO_2$, with 3x3x2 supercells. The k-mesh of size 5 x 5 x 5 in the first Brillouin zone has been used. The self-consistent calculations were considered to be converged, when the total energy of the system is stable within $10^{-3}$mRy. All pure and deficient structures are fully relaxed until the forces per atom declined to less than 0.04 eV/A°, with the energy convergence of $5x10^{-5}$ eV. The optical properties of the two systems, in form of their absorption coefficients, with and without point defects in $SrTiO_3$ and $TiO_2$ have been obtained through complex dielectric functions using the formalism of Ehrenreich and Cohen[38]. The imaginary part of the dielectric function is given as

$$\varepsilon_2(\omega) = \frac{e^2\hbar}{\pi m^2 \omega^2} \sum_{v,c} \int_{BZ} |D_{cv}(k)|^2 \delta[\omega_{cv}(k) - \omega] d^3k \qquad (1)$$

where, $D_{cv} = \langle u_{ck}|e\nabla|u_{vk}\rangle$ matrix elements[39], $\omega_{cv}(k) = E_{ck} - E_{vk}$ is the excitation energy, e is the polarization vector of the electric field. The complex dielectric function is given by

$$\varepsilon(\omega) = \varepsilon_1(\omega) + \varepsilon_2(\omega), \qquad (2)$$

From which the optical absorption coefficient $\alpha(\omega)$ is computed as

$$\alpha(\omega) = \sqrt{2}\omega \left(\sqrt{\varepsilon_1(\omega)^2 + \varepsilon_2(\omega)^2} - \varepsilon_1(\omega)\right)^{1/2}, \qquad (3)$$

**Calculation of defect formation energy**

The formation energies of the neutral vacancies $V^\alpha$ ($\alpha$ = Ti, Sr, O both in $SrTiO_3$ and $TiO_2$) were calculated using the supercell approach[15,19,40,41]. For defect $d$ in charge state $q$, the formation energy $\Delta H(d^q)$ is expressed by-

$$\Delta H(d^q) = E_{tot}(d^q) - E_{tot}(bulk) - \sum \Delta n_i \mu_i + q(E_v + E_F + \Delta V), \qquad (4)$$

Where $E_{tot}(d^q)$ and $E_{tot}(bulk)$ are the total energies of the $SrTiO_3$ supercell with and without defect d, respectively, and $E_F$ is the Fermi energy of the system measured from the VBM. Moreover, $\Delta n_i$ is the number of atoms of species $i$ ($i$ = Sr, Ti or O) that are removed from ($\Delta n_i < 0$) and/or added to ($\Delta n_i > 0$) the supercell to create the defect. $\mu_i$ is the corresponding chemical potential.

**Results and Discussions:**

Figure-1 shows the representative Raman spectra of self-doped $SrTiO_3$ and $TiO_2$. Pure $TiO_2$ and $SrTiO_3$ confirm the structural phase purity of the prepared samples. As disorder increases (i.e. Sr, Ti or O defects), the intensity and FWHM of the vibrational mode near 445 cm$^{-1}$ in $TiO_2$ and 540 cm$^{-1}$ in $SrTiO_3$ increases.

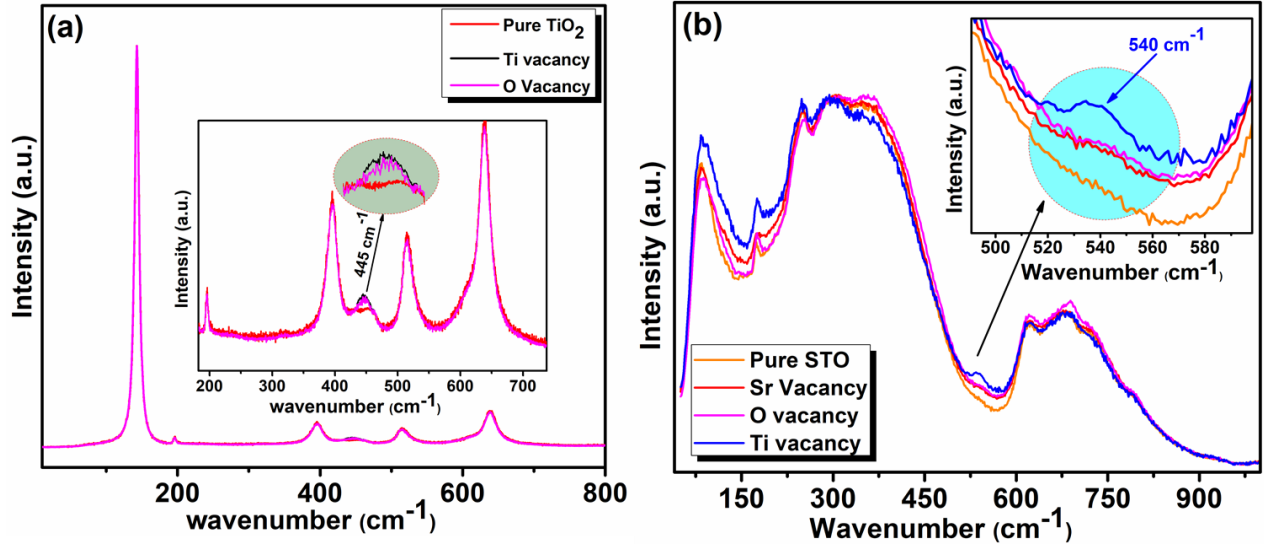

**Figure-1:** *Raman Spectra of prepared pure and defective samples of (a) TiO$_2$ and (b) SrTiO$_3$.*

These pure and deficient samples have been used for optical absorption measurements to see energy states in between valance band maxima (VBM) and conduction band minima (CBM). Mostly oxygen vacancies are inherent in the prepared samples which are not identified through Raman spectra or any other structural characterization. But, native point defects such as these not only cause structural relaxations,[42,43] but also lead to changes to the electronic structure[44,45] of the crystal by inducing defect levels which can be detected in optical spectroscopy. Therefore, theoretical calculations have been performed to calculate electronic structure and optical absorption spectra of these materials with the oxygen vacancies. Before starting the density of states and optical calculations, it is also important to know about formation energies of pure and defected samples. From Table-1 it is obvious that the formation energies of oxygen vacancies are more favorable than other possible vacancies. This result of ours is in full agreement with the previous experimental and theoretical studies in which it was shown that the oxygen vacancy is the fundamental and most common intrinsic defect in perovskite oxides[46–49].

**Table-1:** *Formation energy of pure and self-doped SrTiO$_3$ and TiO$_2$*

| Sample | Formation Energy (eV) |
|---|---|
| TiO$_2$ | -0.25623 |
| TiO$_{2-\delta}$ | 0.01931 |
| Ti$_{1-x}$O$_2$ | 0.02435 |
| SrTiO$_3$ | -0.57562 |
| $Sr_{1-x}TiO_3$ | 0.01327 |
| $SrTi_{1-x}O_3$ | 0.02529 |
| $SrTiO_{3-\delta}$ | 0.01293 |

To further probe this point, the calculated formation energies for the perfect super cells of $TiO_2$ and $SrTiO_3$, and those containing various kinds of neutral native point defects; have been plotted as functions of the Fermi energy ($E_f$), in Figure 2. In the calculations, the zero of the energy was set at the top of the valence band. We have studied the structural relaxations and defect levels induced by various native point defects of selected semiconducting oxides in O-rich (Ti-poor) and O-poor (Ti-rich). By comparing the formation energies, it is concluded that O vacancies are most stable defect in semiconducting oxides like $TiO_2$ and $SrTiO_3$ in comparison to other native defects (see Figure 2 (a-d)).

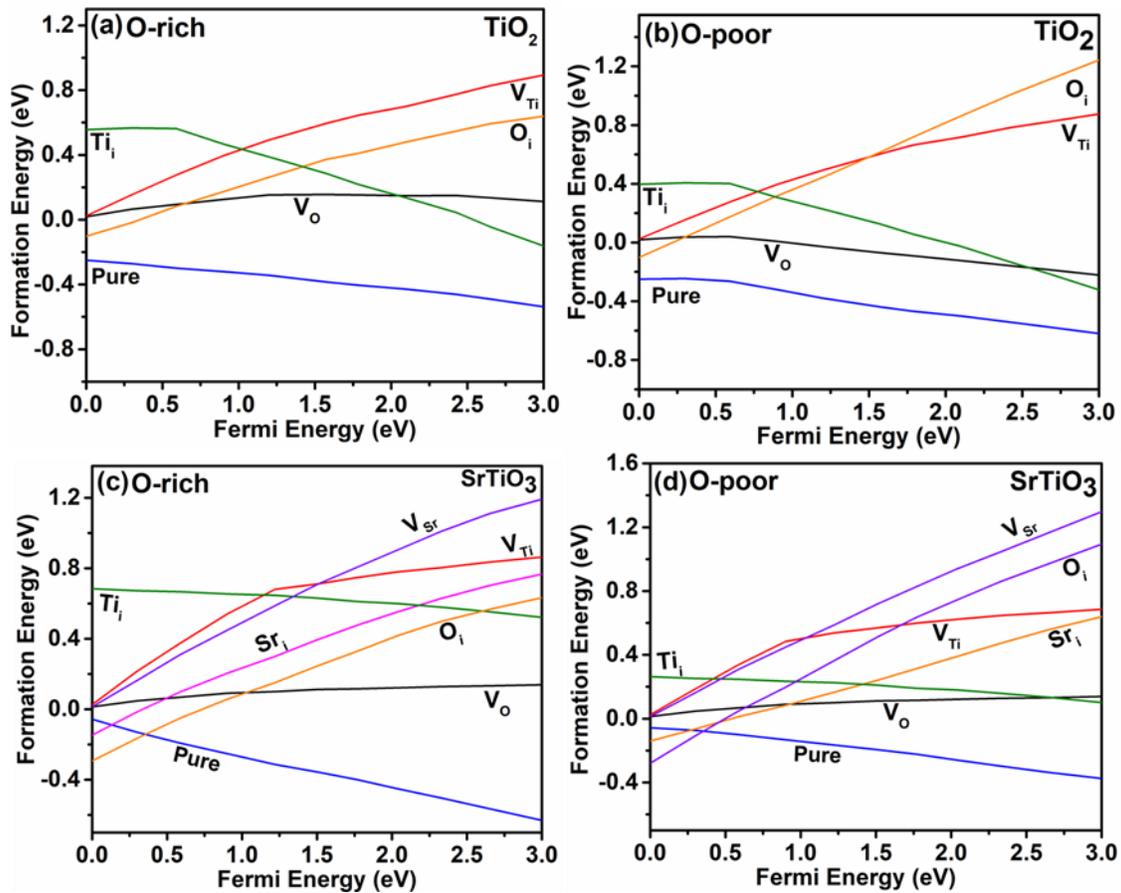

Figure-2: *Defect formation energies along with Fermi energy for (a) O-rich $TiO_2$, (b) O-poor $TiO_2$ (c) O-rich $SrTiO_3$ and (d) O-poor $SrTiO_3$.*

Figure-3 shows the calculated partial density of states (PDOS) of pure $TiO_2$[46], and one can note a complete absence of defect states in between CBM and VBM. The inset of Figure-3 shows the same quantities for (i) Ti-vacancy, and (ii) O-vacancy of $TiO_2$. It is clear from both the insets that due to the vacancy in the material, defect states arise in between CBM and VBM.

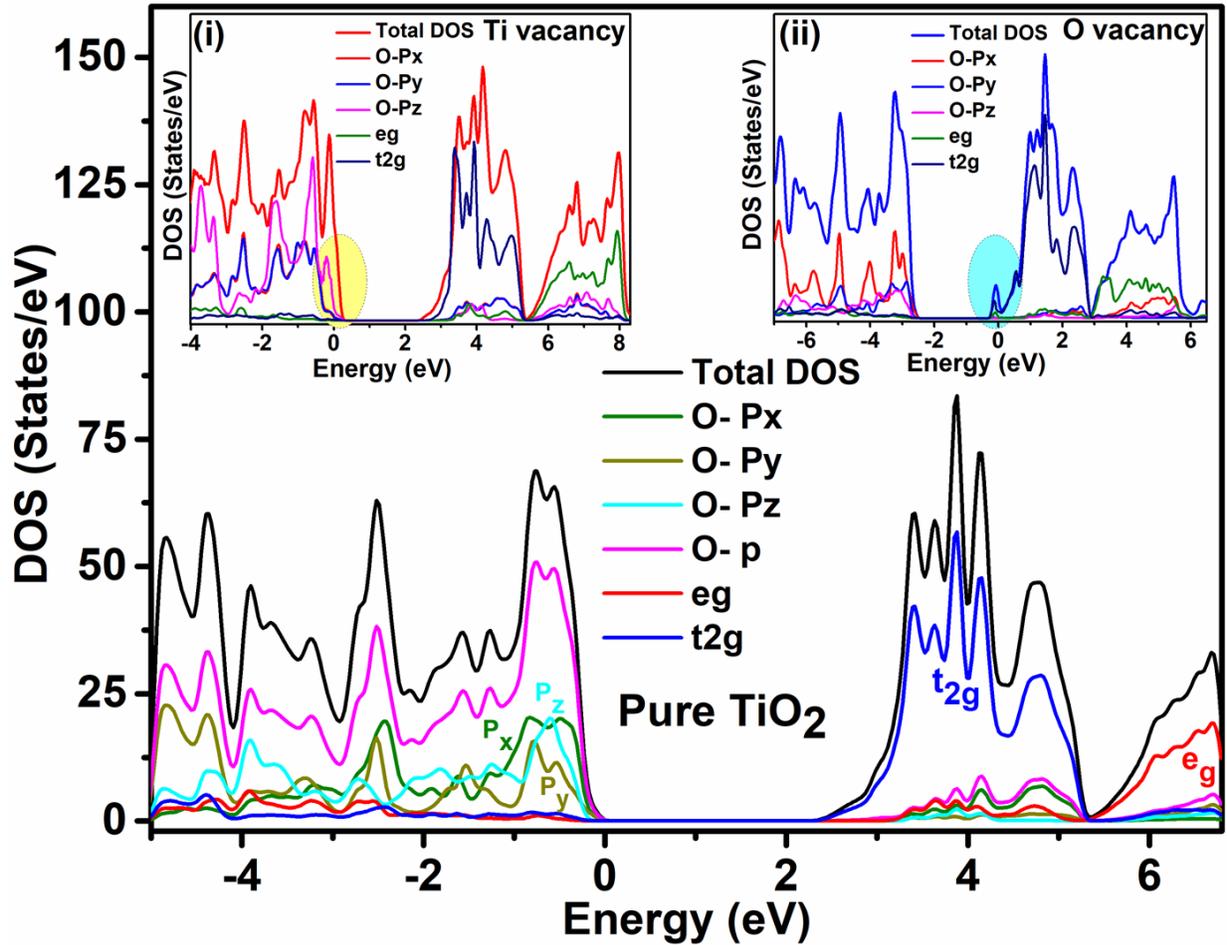

**Figure 3:** *Calculated PDOS for pure $TiO_2$ while in insets we plot the PDOS of the material with: (i) titanium vacancy, and (ii) oxygen vacancy.*

Similarly for $SrTiO_3$, in Fig. 4 we have plotted TDOS and PDOS, and found that a complete absence of defect states in case of pure $SrTiO_3$ (red color), while in case of O, Sr and Ti deficiencies, defect states appear in between CBM and VBM (shown in blue, green and black colors, respectively). It is clear that for pure $TiO_2$, states near Fermi level are dominated by the O-2p and Ti-3d orbitals, signifying hybridization between the two sets of orbitals leading to $t_{2g}$ and $e_g$ states. In case of deficient system, the coexistence of two different oxidation state of Ti

(i.e. $Ti^{+4}$ and $Ti^{+3}$), i.e. mixed valence, has been observed. Similarly, for $SrTiO_3$ also we have observed that for the pure system, CBM and VBM are mainly due to hybridized orbitals consisting of O-2p and Ti-3d states[50,51]. While, for deficient $SrTiO_3$ (see inset-(i) of figure-4) exhibits mixed oxidation states of titanium, i.e., $Ti^{+4}$ and $Ti^{+3}$. Bader[52–54], established an instinctive way of finding charges localized on individual atoms that are part of an extended system (molecule, crystal, etc.), based on the electronic charge density present on those atoms. Hence, to further confirm the variable oxidation states of Ti atoms, we have performed Bader charge analysis[52–54] for $SrTiO_3$ (see Supporting Information) which indicates a clear mixing of transition states $Ti^{+4}$ to $Ti^{+3}$.

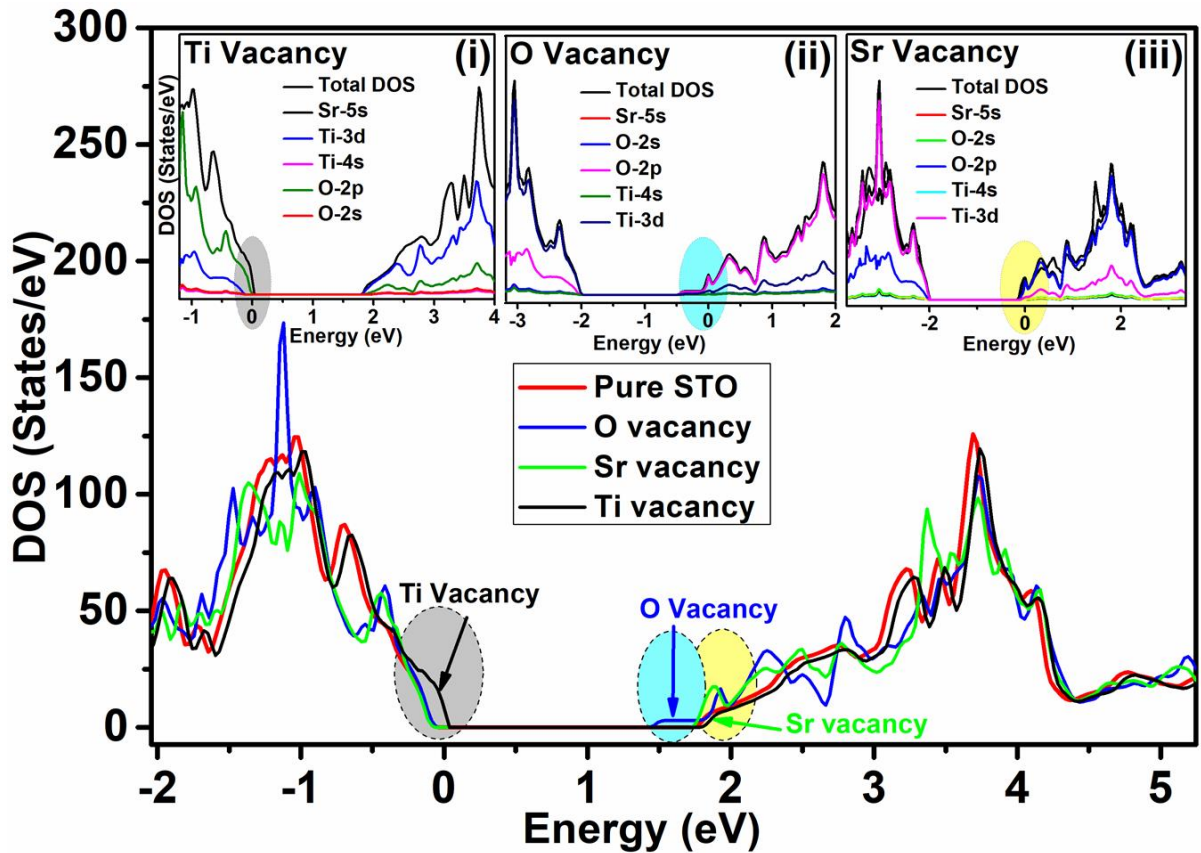

**Figure 4:** *TDOS of pure and self-deficient $SrTiO_3$. Insets show PDOS for deficient $SrTiO_3$ with: (i): Ti vacancies, (ii) O vacancies, and (iii) Sr vacancies*

Because of the presence of defect states in the band gap, it is highly likely that we will see their signatures in the optical spectroscopy performed of these systems Therefore, we have computed the optical absorption spectra of pristine as well as defective $TiO_2$ and $SrTiO_3$, and our results are presented in Fig. 5. These calculations were carried out using the GGA+U approach [36,37,55], with

U=5eV for $TiO_2$ and U=4eV for $SrTiO_3$. The values of U were chosen in such a way that the theoretical band gaps of the pristine systems match their experimental values. First, the optical absorption spectra of pure materials were computed using 3x2x2 geometry-optimized supercells. The spectra of defective systems were calculated after introducing a single vacancy in the supercell at Sr, Ti, or O sites, followed by reoptimization of the geometry.

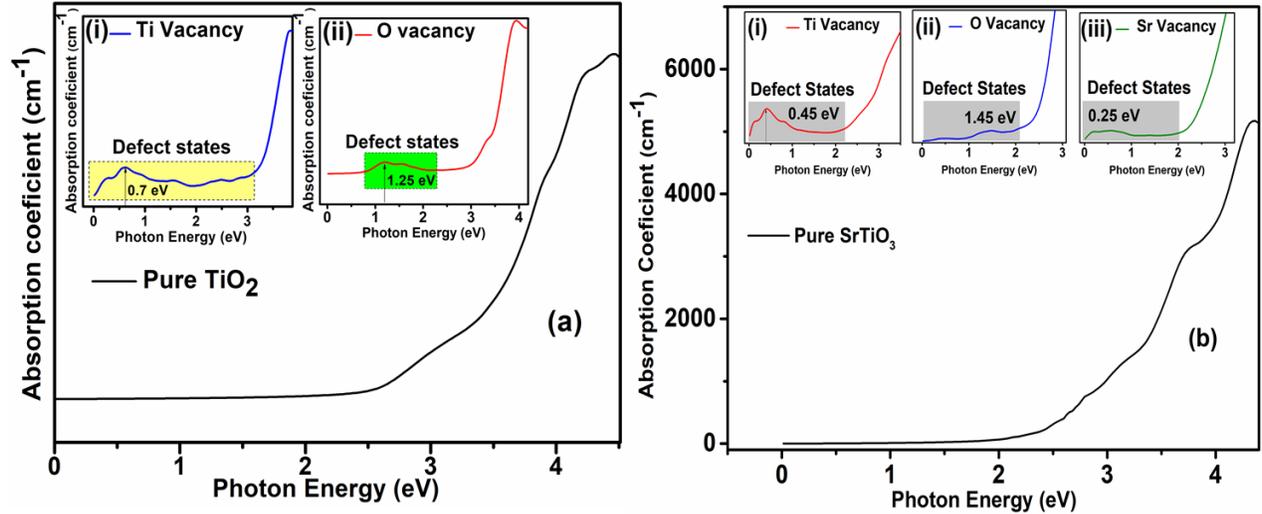

**Figure 5:** *Calculated optical absorption spectra of pristine and defective (a) $TiO_2$, and (b) $SrTiO_3$. Main plots show the spectra of pure systems, while insets show the additional features in the spectra due to vacancy induced defect states.*

On examining Fig. 5, we observe the following trends: (a) for the pure materials, the absorption begins at the gap energy, with no features in the midgap region, (b) in case of defective $TiO_2$ (see figure-5 (a) insets), Ti vacancy leads to a midgap absorption feature at 0.7 eV, while the O vacancy causes a small peak at 1.25 eV, (c) for the defective $SrTiO_3$ (figure-5 (b) insets), Sr, Ti, and O vacancies lead to midgap absorption peaks at 0.25 eV, 0.45 eV, and 1.45 eV, respectively. When we compare the vacancy -induced states in the two compounds, we find that the midgap feature due to Ti vacancy is blue shifted in $TiO_2$ as compared to $SrTiO_3$, while opposite is the case for the O vacancy states in the two materials. Figure-6(a) shows the schematics of the devices obtained by depositing thin films of the semiconducting oxides on the ITO substrate, along with the mechanism for production of photocurrent from the surface of sample, in presence of an applied field. The photocurrents have been measured in the air atmosphere by switching the UV light (wavelength = 350 nm) ''ON'' and ''OFF'' for 20 minutes each, as shown in Figure-6 (b) and (c). Figure-6 (b) plots the measured photocurrents of the prepared devices, both for the pure and defective samples of $TiO_2$, with respect to time. It is clear from figure-6(b) that the device made up of $TiO_2$ sample with vacancies exhibits significant increase in photocurrent as compared to the pure sample. Similarly figure-6(c) shows photocurrent of prepared device of pure and self-deficient $SrTiO_3$ with respect to time. We note that in this case that the maximum photocurrent is observed for Ti-deficient samples, while the minimum one is observed for the sample with Sr vacancies.

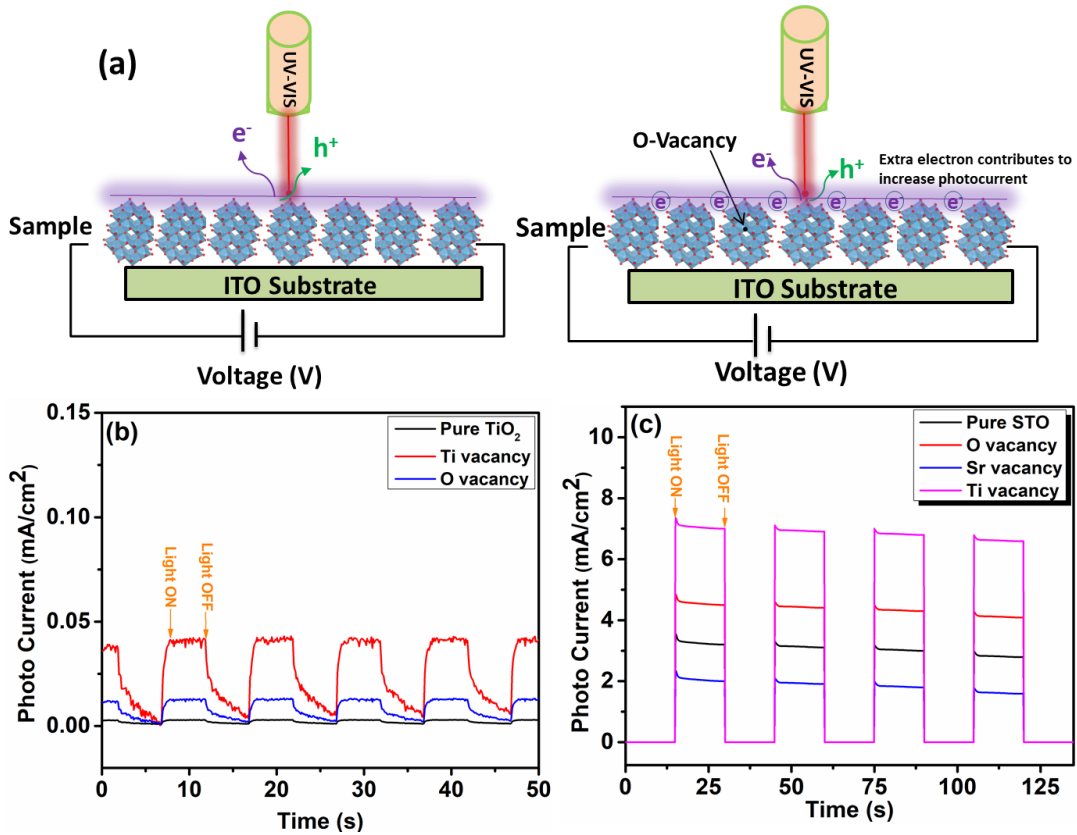

**Figure-6:** *(a) Schematics of the devices obtained by depositing semiconducting oxides on the ITO substrate, along with the mechanism for producing photocurrent in the presence of an applied field. Measured photocurrent values of pure and deficient (b) $TiO_2$, (c) $SrTiO_3$*

We also measured the I-V characteristics of the pure and defective thin films and the plots are presented in figures-7(a) and (b). From the figures it is clearly observed that with the increase in the voltage with the UV light "ON", the photocurrent also increases. When UV light is "ON", current of thin film based devices increases sharply for samples with defects, as compared to the pure ones. From Figs. 7(a) and (b), we conclude that for the case of samples with vacancies, the value of photocurrent enhances with the applied voltage both due to the band edge absorption, and the production of photo carriers from vacancy sites of as prepared thin films. These results indicate that the vacancy-based thin film devices[56] are highly UV sensitive and hence can be used as UV photo-detectors (except for A-type deficiency in $ABO_3$ perovskite). Therefore, we conclude that by creating the vacancies in a controlled manner, we can manipulate the photocurrent in semiconducting devices. Furthermore, by measuring the I-V characteristics of a device, we can identify the nature of vacancies in the material [57–59].

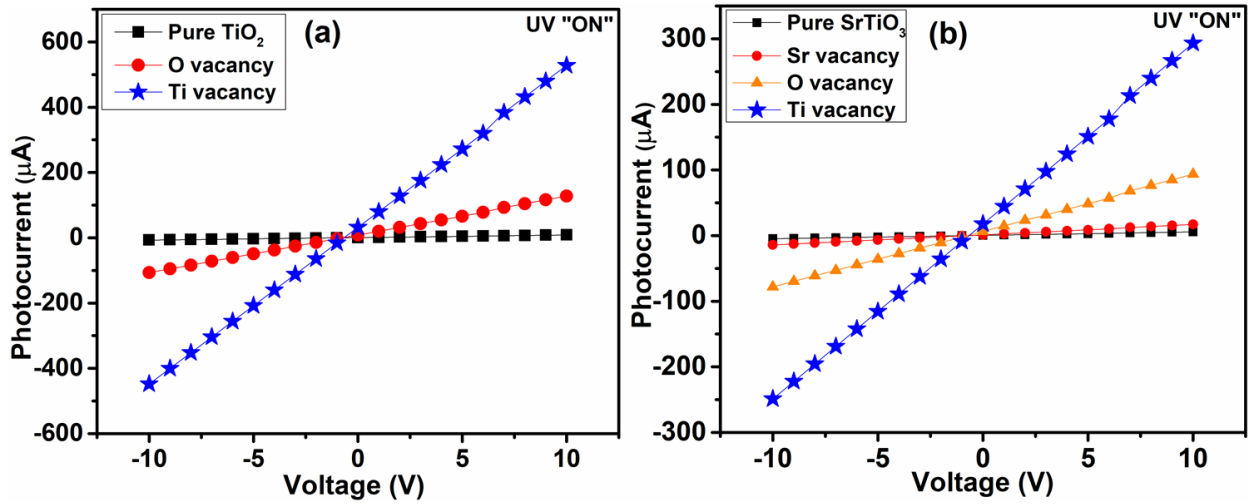

**Figure-7:** *Current-voltage characteristics of (a) pure and vacancy induced $TiO_2$ and (b) pure and vacancy induced $SrTiO_3$ in presence of UV light.*

**Summary and conclusion**

In summary we have shown that vacancies at Sr, Ti and O sites in $SrTiO_3$ and $TiO_2$ lead to the creation of midgap defect states which can be detected both in the optical absorption spectra, as well as the I-V characteristics of the samples. The first-principles calculations on the pure and self-doped $SrTiO_3$ samples suggest that the extra feature in the absorption spectra are essentially due to the vacancies in $SrTiO_3$ and $TiO_2$. Our experimental measurements reveal that vacancies lead to significant changes in the photocurrent as well as I-V characteristics of the samples. This observation can be used to develop novel optoelectronic devices based on self-deficient semiconducting oxides.

## Acknowledgments

One of the authors (SP) acknowledges the Homi Bhabha Research Cum Teaching Fellowship (A.K.T.U.), Lucknow, India for providing financial support Through Teaching Assistantship.

## CRediT authorship contribution statement

**Shashi Pandey:** Conceptualization, Data curation, Investigation, Methodology, Formal analysis, Writing – original draft. **Alok Shukla:** Resources, Software, Supervision, Validation, Writing – review & editing. **Anurag Tripathi:** Conceptualization, Funding acquisition, Resources, Software, Supervision, Validation, Writing – review & editing.

## Compliance with ethical standards:

Conflict of interest: The authors declare that they do not have any conflict of interest.

## Data availability statement –

The raw/processed data required to reproduce these findings cannot be shared at this time as the data also forms part of an ongoing study.

## References

[1] C.S. Koonce, M.L. Cohen, J.F. Schooley, W.R. Hosler, and E.R. Pfeiffer, Phys. Rev. **163**, 380 (1967).
[2] N.A. Pertsev, A.K. Tagantsev, and N. Setter, Phys. Rev. B **61**, R825 (2000).
[3] O. Ivanov, E. Danshina, Y. Tuchina, and V. Sirota, Phys. Status Solidi B **248**, 1006 (2011).
[4] J.M.D. Coey, M. Venkatesan, and P. Stamenov, J. Phys. Condens. Matter **28**, 485001 (2016).
[5] S. Zhang, D. Guo, M. Wang, M.S. Javed, and C. Hu, Appl. Surf. Sci. **335**, 115 (2015).
[6] K. Janicka, J.P. Velev, and E.Y. Tsymbal, J. Appl. Phys. **103**, 07B508 (2008).
[7] N. Baskaran, A. Ghule, C. Bhongale, R. Murugan, and H. Chang, J. Appl. Phys. **91**, 10038 (2002).
[8] K. Kinoshita and A. Yamaji, J. Appl. Phys. **47**, 371 (1976).
[9] K. Xie, N. Umezawa, N. Zhang, P. Reunchan, Y. Zhang, and J. Ye, Energy Environ. Sci. **4**, 4211 (2011).
[10] R. Courths, B. Cord, and H. Saalfeld, Solid State Commun. **70**, 1047 (1989).
[11] X. Bai, L. Wang, R. Zong, Y. Lv, Y. Sun, and Y. Zhu, Langmuir **29**, 3097 (2013).
[12] V.E. Alexandrov, E.A. Kotomin, J. Maier, and R.A. Evarestov, Eur. Phys. J. B **72**, 53 (2009).
[13] R.S. Dima, L. Phuthu, N.E. Maluta, J.K. Kirui, and R.R. Maphanga, Materials **14**, 3918 (2021).
[14] O.A. Dicks, A.L. Shluger, P.V. Sushko, and P.B. Littlewood, Phys. Rev. B **93**, 134114 (2016).
[15] M. Choi, F. Oba, and I. Tanaka, Appl. Phys. Lett. **98**, 172901 (2011).
[16] M.W. Khan, S. Husain, M.A.M. Khan, M. Gupta, R. Kumar, and J.P. Srivastava, Philos. Mag. **90**, 3069 (2010).

[17] Q. Liu, Y. Zhou, J. Kou, X. Chen, Z. Tian, J. Gao, S. Yan, and Z. Zou, J. Am. Chem. Soc. **132**, 14385 (2010).
[18] Z. Yi, J. Ye, N. Kikugawa, T. Kako, S. Ouyang, H. Stuart-Williams, H. Yang, J. Cao, W. Luo, Z. Li, Y. Liu, and R.L. Withers, Nat. Mater. **9**, 559 (2010).
[19] Z. Chao, W. Chun-Lei, L. Ji-Chao, and Y. Kun, Chin. Phys. **16**, 1422 (2007).
[20] D. Kan, T. Terashima, R. Kanda, A. Masuno, K. Tanaka, S. Chu, H. Kan, A. Ishizumi, Y. Kanemitsu, Y. Shimakawa, and M. Takano, Nat. Mater. **4**, 816 (2005).
[21] J. Yan, G. Wu, N. Guan, L. Li, Z. Li, and X. Cao, Phys. Chem. Chem. Phys. **15**, 10978 (2013).
[22] B. Modak, K. Srinivasu, and S. K. Ghosh, Phys. Chem. Chem. Phys. **16**, 24527 (2014).
[23] H. Tan, Z. Zhao, W. Zhu, E.N. Coker, B. Li, M. Zheng, W. Yu, H. Fan, and Z. Sun, ACS Appl. Mater. Interfaces **6**, 19184 (2014).
[24] N.C. Bristowe, P.B. Littlewood, and E. Artacho, Phys. Rev. B **83**, 205405 (2011).
[25] A. Lopez-Bezanilla, P. Ganesh, and P.B. Littlewood, Phys. Rev. B **92**, 115112 (2015).
[26] N.C. Bristowe, P. Ghosez, P.B. Littlewood, and E. Artacho, J. Phys. Condens. Matter **26**, 143201 (2014).
[27] S. John, C. Soukoulis, M.H. Cohen, and E.N. Economou, Phys. Rev. Lett. **57**, 1777 (1986).
[28] E.A. Davis and N.F. Mott, Philos. Mag. **22**, 0903 (1970).
[29] G. Gouadec and P. Colomban, Prog. Cryst. Growth Charact. Mater. **53**, 1 (2007).
[30] (n.d.).
[31] Y. Li, W.G. Schmidt, and S. Sanna, Phys. Rev. B **91**, 174106 (2015).
[32] P. Hohenberg and W. Kohn, Phys. Rev. **136**, B864 (1964).
[33] W. Kohn and L.J. Sham, Phys. Rev. **140**, A1133 (1965).
[34] G Kresse, J Hafner, J Phys Condens Matter 6 8245 (1994).
[35] G. Kresse, J.Furthmuller, Phys Rev B 54 11169 (1996).
[36] K. Yang, Y. Dai, B. Huang, and Y.P. Feng, J. Phys. Appl. Phys. **47**, 275101 (2014).
[37] A. Rubio-Ponce, A. Conde-Gallardo, and D. Olguín, Phys. Rev. B **78**, 035107 (2008).
[38] H. Ehrenreich and M.H. Cohen, Phys. Rev. **115**, 786 (1959).
[39] R. Khenata, M. Sahnoun, H. Baltache, M. Rérat, A.H. Rashek, N. Illes, and B. Bouhafs, Solid State Commun. **136**, 120 (2005).
[40] Z. Hu and H. Metiu, J. Phys. Chem. C **115**, 17898 (2011).
[41] A. Janotti, J.B. Varley, P. Rinke, N. Umezawa, G. Kresse, and C.G. Van de Walle, Phys. Rev. B **81**, 085212 (2010).
[42] G. Rahman, N.U. Din, V.M. García-Suárez, and E. Kan, Phys. Rev. B **87**, 205205 (2013).
[43] S.B. Zhang, S.-H. Wei, and A. Zunger, Phys. Rev. B **63**, 075205 (2001).
[44] S. Na-Phattalung, M.F. Smith, K. Kim, M.-H. Du, S.-H. Wei, S.B. Zhang, and S. Limpijumnong, Phys. Rev. B **73**, 125205 (2006).
[45] F. Oba, M. Choi, A. Togo, A. Seko, and I. Tanaka, J. Phys.: Condens. Matter **22**, 384211 (2010).
[46] M. Sotoudeh, S.J. Hashemifar, M. Abbasnejad, and M.R. Mohammadizadeh, AIP Adv. **4**, 027129 (2014).
[47] H. Li, Y. Guo, and J. Robertson, J. Phys. Chem. C **119**, 18160 (2015).
[48] F. Wang, C. Di Valentin, and G. Pacchioni, Phys. Rev. B **84**, 073103 (2011).
[49] M. Sun, W. Li, B. Zhang, G. Cheng, B. Lan, F. Ye, Y. Zheng, X. Cheng, and L. Yu, Chem. Eng. J. **331**, 626 (2018).
[50] P. Singh, S. Ghosh, V. Mishra, S. Barman, S. Roy Barman, A. Singh, S. Kumar, Z. Li, U. Kentsch, and P. Srivastava, J. Magn. Magn. Mater. **523**, 167630 (2021).
[51] P. Singh, V. Mishra, S. Barman, M. Balal, S.R. Barman, A. Singh, S. Kumar, R. Ramachandran, P. Srivastava, and S. Ghosh, J. Alloys Compd. **889**, 161663 (2021).
[52] R.F.W. Bader and H. Essén, J. Chem. Phys. **80**, 1943 (1984).
[53] D.A. Carballo-Córdova, M.T. Ochoa-Lara, S.F. Olive-Méndez, and F. Espinosa-Magaña, Philos. Mag. **99**, 181 (2019).

[54] G. Henkelman, A. Arnaldsson, and H. Jónsson, Comput. Mater. Sci. **36**, 354 (2006).
[55] M.H. Samat, M.F.M. Taib, N.K. Jaafar, O.H. Hassan, M.Z.A. Yahya, and A.M.M. Ali, AIP Conf. Proc. **2030**, 020058 (2018).
[56] D. Maarisetty and S.S. Baral, Ceram. Int. **45**, 22253 (2019).
[57] J. Ravichandran, W. Siemons, H. Heijmerikx, M. Huijben, A. Majumdar, and R. Ramesh, Chem. Mater. **22**, 3983 (2010).
[58] W.-J. Yin, S.-H. Wei, M.M. Al-Jassim, and Y. Yan, Phys. Rev. B **85**, 201201 (2012).

## Supporting Information

**Table-2:** *Bader analysis of pure and self-doped $SrTiO_3$*

| No. | Atoms | SrTiO$_3$ | Sr vacancy | Ti vacancy | O vacancy |
|---|---|---|---|---|---|
| 1 | Sr1 | 1.954 | 1.923 | 1.92 | 1.918 |
| 2 | Sr2 | 1.954 | 1.923 | 1.92 | 1.918 |
| 3 | Sr3 | 1.968 | 1.945 | 1.941 | 1.932 |
| 4 | Sr4 | 1.924 | 1.931 | 1.929 | 1.911 |
| 5 | Sr5 | 1.947 | 1.953 | 1.943 | 1.932 |
| 6 | Sr6 | 1.956 | 1.962 | 1.951 | 1.934 |
| 7 | Sr7 | 1.942 | 1.926 | 1.932 | 1.921 |
| 8 | Sr8 | 1.978 | 1.985 | 1.977 | 1.962 |
| 9 | Sr9 | 1.975 | 1.978 | 1.976 | 1.952 |
| 10 | Sr10 | 1.947 | 1.949 | 1.954 | 1.941 |
| 11 | Sr11 | 1.958 | 1.968 | 1.952 | 1.945 |
| 12 | Sr12 | 1.957 | 1.961 | 1.951 | 1.943 |
| 13 | Sr13 | 1.914 | 1.92 | 1.918 | 1.905 |
| 14 | Sr14 | 1.991 | 1.983 | 1.981 | 1.971 |
| 15 | Sr15 | 1.925 | 1.929 | 1.921 | 1.917 |
| 16 | Sr16 | 1.969 | 1.962 | 1.965 | 1.943 |
| 17 | Sr17 | 1.959 | 1.951 | 1.953 | 1.941 |
| 18 | Sr18 | 1.929 | 1.932 | 1.921 | 1.916 |
| 19 | Ti1 | 3.274 | 3.312 | 3.521 | 3.772 |
| 20 | Ti2 | 3.375 | 3.453 | 3.516 | 3.575 |
| 21 | Ti3 | 3.264 | 3.376 | 3.512 | 3.668 |
| 22 | Ti4 | 3.212 | 3.289 | 3.375 | 3.452 |
| 23 | Ti5 | 3.157 | 3.215 | 3.331 | 3.423 |
| 24 | Ti6 | 3.521 | 3.593 | 3.671 | 3.721 |

| 25 | Ti7 | 3.268 | 3.326 | 3.381 | 3.423 |
| --- | --- | --- | --- | --- | --- |
| 26 | Ti8 | 3.412 | 3.478 | 3.531 | 3.598 |
| 27 | Ti9 | 3.475 | 3.521 | 3.532 | 3.638 |
| 28 | Ti10 | 3.372 | 3.432 | 3.416 | 3.467 |
| 29 | Ti11 | 3.174 | 3.252 | 3.233 | 3.338 |
| 30 | Ti12 | 3.371 | 3.411 | 3.442 | 3.51 |
| 31 | Ti13 | 3.521 | 3.598 | 3.598 | 3.617 |
| 32 | Ti14 | 3.488 | 3.542 | 3.567 | 3.671 |
| 33 | Ti15 | 3.442 | 3.499 | 3.587 | 3.591 |
| 34 | Ti16 | 3.327 | 3.394 | 3.412 | 3.487 |
| 35 | Ti17 | 3.716 | 3.786 | 3.796 | 3.854 |
| 36 | Ti18 | 3.627 | 3.687 | 3.698 | 3.678 |
| 37 | O 1 | -1.86 | -1.869 | -1.894 | -1.899 |
| 38 | O 2 | -1.923 | -1.973 | -1.942 | -1.991 |
| 39 | O 3 | -1.823 | -1.843 | -1.862 | -1.876 |
| 40 | O 4 | -1.821 | -1.834 | -1.867 | -1.872 |
| 41 | O 5 | -1.853 | -1.866 | -1.883 | -1.894 |
| 42 | O 6 | -1.821 | -1.852 | -1.871 | -1.882 |
| 43 | O 7 | -1.732 | -1.751 | -1.776 | -1.802 |
| 44 | O 8 | -1.782 | -1.791 | -1.812 | -1.819 |
| 45 | O 9 | -1.754 | -1.788 | -1.827 | -1.839 |
| 46 | O 10 | -1.921 | -1.923 | -1.963 | -1.983 |
| 47 | O 11 | -1.952 | -1.974 | -1.955 | -1.986 |
| 48 | O 12 | -1.751 | -1.769 | -1.784 | -1.854 |
| 49 | O 13 | -1.854 | -1.878 | -1.896 | -1.965 |
| 50 | O 14 | -1.832 | -1.864 | -1.921 | -1.942 |
| 51 | O 15 | -1.568 | -1.573 | -1.662 | -1.712 |
| 52 | O 16 | -1.921 | -1.918 | -1.934 | -1.953 |
| 53 | O 17 | -1.728 | -1.753 | -1.792 | -1.805 |
| 54 | O18 | -1.561 | -1.587 | -1.714 | -1.765 |